\documentclass[11pt,a4paper]{article}
\begin{document}

\begin{center}
{\Large Shape/Phase Transitions and Critical Point Symmetries in Atomic Nuclei}

\bigskip \bigskip

\small

Dennis Bonatsos

\bigskip

\footnotesize

{\em Institute of Nuclear Physics, National Centre for Scientific Research 
``Demokritos'', GR-15310 Aghia Paraskevi, Attiki, Greece,
E-mail: bonat@inp.demokritos.gr }

\bigskip

\end{center}

\bigskip

\footnotesize

{\em Abstract.\/} 

Shape/phase transitions in atomic nuclei have first been discovered 
in the framework of the Interacting Boson Approximation (IBA) model. 
Critical point symmetries appropriate for nuclei at the transition points 
have been introduced as special solutions of the Bohr Hamiltonian, 
stirring the introduction of additional new solutions describing wide 
ranges of nuclei. A short review of these recent developments will be 
attempted. 

\bigskip


\normalsize


\section{INTRODUCTION} 

Atomic nuclei are known to exhibit changes of their energy levels and 
electromagnetic transition rates among them when the number of protons 
and/or neutrons is modified, resulting in shape phase transitions from one 
kind of collective behaviour to another. These transitions are not phase 
transitions of the usual thermodynamic type. They are quantum phase 
transitions \cite{IacIJMPA}
(initially called ground state phase transitions \cite{Deans}), occurring 
in Hamiltonians of the type 
$ H = c(H_1 + g H_2)$,
where $c$ is a scale factor, $g$ is the control parameter, and $H_1$, $H_2$ describe two different phases of the system. The expectation value of a 
suitably chosen operator, characterizing the state of the system, is used as
the order parameter.

In the framework of the Interacting Boson Model \cite{IA}, which describes 
nuclear structure of even--even nuclei within the U(6) symmetry, possessing 
the U(5), SU(3), and O(6) limiting dynamical symmetries, appropriate for 
vibrational, axially deformed, and $\gamma$-unstable nuclei respectively, 
shape phase transitions have been studied 25 years ago \cite{Deans}
using the classical limit of the 
model \cite{GK,GK2,DSI,vanRoos}, 
pointing out that there is (in the usual Ehrenfest classification) a second order shape phase transition between U(5) and O(6), a first order 
shape phase transition between U(5) and SU(3), and no shape phase transition 
between O(6) and SU(3). It is instructive to place \cite{IacIJMPA}
these shape phase transitions 
on the symmetry triangle of the IBM \cite{Casten}, at the three corners of 
which the three limiting symmetries of the IBM appear. 

More recently it has been realized \cite{IacE5,IacX5} that the properties 
of nuclei lying at the critical point of a shape phase transition can be 
described by appropriate special solutions of the Bohr Hamiltonian \cite{Bohr},
labelled as critical point symmetries. The E(5) critical point symmetry \cite{IacE5} has been found to correspond to the second order critical point between U(5) and O(6), while the X(5) critical point symmetry \cite{IacX5}
has been found to correspond to the first order transition between 
U(5) and SU(3).  

The introduction of the critical point symmetries E(5) \cite{IacE5} and X(5) 
\cite{IacX5} has triggered 
much work on special solutions of the Bohr Hamiltonian, corresponding to 
different physical situations. Several of these solutions will be mentioned 
here, together with experimental examples appropriate for each case. 

Two recent developments should be mentioned here. Considering interacting 
boson models with two types of bosons (one scalar, one non-scalar) with 
U($n$) symmetry and the relevant classical descriptions in terms of $n-1$
variables, it has been proved that both first and second order phase 
transitions occur only for $n=6$, 10, 14, \dots, if rotational invariance 
is assumed, while in the rest of the cases only second order transitions 
occur \cite{CI}. Furthermore, the study of excited state phase transitions
(in contrast to the ground state phase transitions mentioned above) 
has started in the framework of several many-body models \cite{CCI}.  

\section{Shape phase transitions in the Interacting Boson Model}

In the framework of the Interacting Boson Model \cite{IA}, 
nuclear structure of even--even nuclei is described 
within the U(6) symmetry, possessing 
the U(5), SU(3), and O(6) limiting dynamical symmetries, appropriate for 
vibrational, axially deformed, and $\gamma$-unstable nuclei respectively.

A form of the IBM Hamiltonian appropriate for the study of shape phase 
transitions reads \cite{Zamfir66,Werner} 
\begin{equation}\label{eq:e1a} 
H(\zeta,\chi) = c \left[ (1-\zeta) \hat n_d -{\zeta\over 4 N_B}
\hat Q^\chi \cdot \hat Q^\chi\right],
\end{equation}
where $\hat n_d = d^\dagger \cdot \tilde d$ is the number operator 
of $d$-bosons, 
$\hat Q^\chi =(s^\dagger \tilde d + d^\dagger s) +\chi (d^\dagger \tilde
d)^{(2)}$ 
is the quadrupole operator,  
$N_B$ is the number of valence bosons, and $c$ is a
scaling factor. The above Hamiltonian contains two parameters,
$\zeta$ and $\chi$, with the parameter $\zeta$ ranging from 0 to
1, and the parameter $\chi$ ranging from 0 to $-\sqrt{7}/2=-1.32$.
In this parametrization, the U(5) limit corresponds to $\zeta=0$, 
the O(6) limit to $\zeta=1$, $\chi=0$, and the SU(3) limit to 
$\zeta=1$, $\chi=-\sqrt{7}/2$.  

It is instructive to place the three limiting symmetries at the corners 
of a triangle, called the symmetry triangle \cite{Casten} of the IBM.  
With the above parametrization,  the entire symmetry triangle of the
IBM can be described, along with each of the
three dynamical symmetry limits of the IBM. The parameters
$(\zeta,\chi)$ can be plotted in the symmetry triangle by
converting them into polar coordinates \cite{McC6906}
\begin{equation}\label{eq:e2d}
\rho = {\sqrt{3} \zeta \over \sqrt{3} \cos\theta_\chi
-\sin\theta_\chi}, \qquad \theta = {\pi \over 3} +\theta_\chi,
\end{equation}
where $\theta_\chi=(2/\sqrt{7})\chi(\pi/3)$.

Shape phase transitions in the IBM can be studied by considering 
the classical limit of the model, which can be obtained by several methods:

a) The method of coherent states \cite{Gilmore,GK,GK2,DSI,vanRoos}.

b) The method of equations of motion \cite{Hatch}.

c) The method involving a Holstein-Primakoff transformation \cite{KV,KLV}

Using the first method \cite{DSI}, one can obtain the energy functionals $E(\beta,\gamma)$ (in terms of the Bohr variables \cite{Bohr} $\beta$ and $\gamma$), corresponding to each symmetry limit of the IBM, and minimize them 
with respect to $\beta$ and $\gamma$. It turns out that the energy functionals 
of the U(5) and O(6) limits are $\gamma$-independent, possessing a single minimum at $\beta=0$ and at $\beta=1$ respectively, while the energy functional 
of the SU(3) limit does depend on $\gamma$, possessing a sharp minimum at 
$\gamma=0$ (corresponding to prolate axial symmetry) and $\beta =\sqrt{2}$. 
The minima are in agreement with the interpretation of the U(5), O(6) 
and SU(3) limits as representing vibrational (spherical), $\gamma$-unstable,
and prolate deformed nuclei respectively. 
Reversing the sign of the $\chi$ parameter in the quadrupole operator 
the minimum appears at $\gamma=60^{\rm o}$ [corresponding to the oblate axial 
symmetry labelled as $\overline {\rm SU(3)}$] and $\beta =\sqrt{2}$. 

Using the classical limit of the IBM it has been realized 25 years ago \cite{Deans} that there is a second order shape phase transition between 
U(5) and O(6), a first order 
shape phase transition between U(5) and SU(3), and no shape phase transition 
between O(6) and SU(3). The usual Ehrenfest classification is used, in which 
the order of the transition corresponds to the order $n$ of the derivative
$\partial^n E  /\partial \zeta^n$ in which discontinuity appears. 

Using the coherent state formalism of the IBA
\cite{GK,GK2,DSI} one can obtain the scaled total energy,
$E(\beta,\gamma)/(c N_B)$, corresponding to the Hamiltonian of 
Eq. (\ref{eq:e1a}), in the form \cite{IZ}
$$ {\cal E}(\beta,\gamma)= {\beta^2 \over 1+\beta^2} \left[
(1-\zeta)-(\chi^2+1) {\zeta \over 4 N_B}\right] -{5\zeta\over 4
N_B (1+\beta^2)} $$
\begin{equation}\label{eq:e3d}
-{\zeta (N_B-1) \over 4 N_B (1+\beta^2)^2} \left[ 4\beta^2 -4
\sqrt{2\over 7} \chi \beta^3 \cos 3\gamma +{2\over 7} \chi^2
\beta^4\right] ,
\end{equation}
where $\beta$ and $\gamma$ are the two classical coordinates,
related \cite{IA} to the Bohr geometrical variables \cite{Bohr}.

According to the results \cite{Deans} mentioned above, one expects a 
first order transition between U(5) and SU(3), i.e. on the leg 
of the symmetry triangle of the IBM characterized by $\chi=-\sqrt{7}/2=-1.32$,  
and a second order transition between U(5) and O(6), i.e. on the leg 
of the symmetry triangle corresponding to $\chi=0$. 

As a function of $\zeta$, a shape/phase coexistence region
\cite{IZC} begins when a deformed minimum appears in addition to
the spherical minimum (which occurs at $\beta=0$)
and ends when only the deformed minimum
remains. The latter is achieved when ${\cal E}(\beta,\gamma)$
becomes flat at $\beta=0$, fulfilling the condition \cite{Werner}
${\partial^2 {\cal E} \over \partial \beta^2} \vert _{\beta=0}=0$,
which is satisfied for
\begin{equation}\label{eq:e4d}
\zeta^{**}= {4 N_B \over 8 N_B +\chi^2-8}.
\end{equation}

The former, $\zeta^*$, can be derived from the results of Ref.
~\cite{catast}. For $\chi=-\sqrt{7}/2$ this point is given by \cite{Libby1}
\begin{equation}
\zeta^* = {(896\sqrt{2} + 656 R)N_B \over -1144\sqrt{2} +123 R
+(1536\sqrt{2} +164R)N_B }
\end{equation}
\noindent where
\begin{equation}
R= \sqrt{ {35456\over 15129}+{32 \enskip 6^{2/3} \over 41} }
-\sqrt{ {70912\over 15129} -{32 \enskip 6^{2/3} \over 41} +
{3602816 \over 15129 \sqrt{1108+369 \enskip 6^{2/3}} } }
\end{equation}

In between there is a point, $\zeta_{\rm crit}$,  where the two
minima are equal and the first derivative of ${\cal E}_{min}$,
$\partial {\cal E}_{min}/\partial \zeta$, is discontinuous,
indicating a first-order phase transition. For $\chi=-1.32$, i.e. on the 
U(5)-SU(3) leg of the symmetry triangle, this point is
\cite{Fernandes}
\begin{equation}\label{eq:e5}
\zeta_{\rm crit} = {16 N_B \over 34 N_B -27}.
\end{equation}
Expressions for $\zeta^*$ and $\zeta_{crit}$ involving
the parameter $\chi$ can also be deduced using the results of Ref.
~\cite{catast}.

The range of $\zeta$ corresponding to the region of shape/phase
coexistence shrinks with decreasing $\vert \chi \vert$ and
converges to a single point for $\chi=0$, which is the point of a
second-order phase transition between U(5) and O(6), located on
the U(5)--O(6) leg of the symmetry triangle (which is
characterized by $\chi=0$)  at $\zeta= N_B/(2N_B-2)$, as seen from
Eq. (\ref{eq:e4d}). 

For $N_B=10$, which is  a value typical for several nuclei,
it is clear that the left
border of the phase transition region, defined by $\zeta^*$, and
the line defined by $\zeta_{\rm crit}$ nearly coincide. For
$\chi=-1.32$, in particular, one has $\zeta^*=0.507$ and
$\zeta_{\rm crit}=0.511$. Therefore one is entitled to use
$\zeta_{\rm crit}$ as the approximate left border of the phase
transition region.

It is instructive to plot the evolution with $\zeta$ of the IBM total energy 
curves for $\chi=-1.32$, i.e. along the U(5)-SU(3) leg of the IBM symmetry triangle, and for a typical constant value of $N_B$ ($N_B=10$, for example). 
At $\zeta=0$ a single minimum at $\beta=0$ occurs. At $\zeta^*= 0.507$, 
a deformed minimum appears in addition to the spherical one. At 
$\zeta_{\rm crit}=0.511$ the two minima are equal, the total energy curve 
exhibiting a bump between them, which is a hallmark of a first order phase 
transition. At $\zeta^{**}=0.542$ the spherical minimum disappears, thus for 
$\zeta> 0.542$ only a deformed minimum exists. 

It is also instructive to plot the evolution with $\zeta$ of the IBM total 
energy curves for $\chi=0$, i.e. along the U(5)-O(6) leg of the IBM symmetry 
triangle, again for $N_B=10$. For $\zeta=0$ only the spherical minimum at 
$\beta=0$ exists. At $\zeta_{crit}=0.556$ the minimum energy jumps to non-zero 
$\beta$, the bottom of the total energy curve being quite flat, which is a 
hallmark of a second order phase transition. 

\subsection{Prolate to oblate transition} 

It has been argued \cite{Jolie87}
that O(6) can be considered as a critical point in the transition from prolate to oblate deformed shapes, i.e. from SU(3) 
($\chi=-1.32$) to $\overline {\rm SU(3)}$ ($\chi=+1.32$). This situation can be 
depicted in the extended symmetry triangle of the IBM, in which the 
$\overline {\rm SU(3)}$ limit is also included. This argument is based 
on the fact that some observables (Q-invariants \cite{Cline,Kumar})  
when plotted as functions of $\chi$, exhibit turning  
points at $\chi=0$, i.e. at O(6), a special behaviour which has been seen 
for the U(5)-SU(3) and U(5)-O(6) transitions \cite{Werner}.  

In this case the point of second order phase transition between U(5) and O(6)
can be interpreted \cite{Jolie89} as a triple point, lying at the point 
where three different regions (spherical, prolate, and oblate) meet. In particular, this triple point is the junction of the line representing the 
first order phase transition from spherical to deformed shapes, and the line 
corresponding to the first order phase transition between prolate and oblate 
shapes, in accordance to Landau theory of phase transitions \cite{Landau},
which predicts the existence of isolated points of second order phase transitions at the intersections of two or more curves corresponding to 
first order phase transitions. 

A chain of nuclei, each differing from the previous one by two protons or two 
neutrons, has been found \cite{Linnemann}, indicating $^{194}$Pt as lying 
close to the critical point of the prolate to oblate transition. 
Rerativistic mean field (RMF) calculations \cite{FossionI} for the same chain 
of nuclei corroborate this conclusion. 
However, the same RMF calculations \cite{FossionI} in the Pt chain 
of isotopes indicate a transition from prolate to oblate shapes between $^{186}$Pt and $^{188}$Pt, while in the Os chain of isotopes they predict a 
transition from prolate to oblate shapes between $^{192}$Os and $^{194}$Os.  

The prediction that the nucleus $^{186}$Pt is critical is supported by 
several pieces of evidence \cite{FossionI}. The $\beta_1$-bandheads 
(normalized to the energy of the $2_1^+$ state) exhibit a minimum 
at $^{186}$Pt, while the crossover of the (normalized to the energy of 
the $2_1^+$ state) bandheads of the $\beta_1$ and $\gamma_1$ bands 
also occurs at the same nucleus. Furthermore, mapping the Pt isotopic chain 
on the IBM symmetry triangle shows \cite{McCPt} that $^{186}$Pt lies very close to the shape 
phase coexistence region of IBM \cite{IZC,McC6906}.

\section{E(5) AND RELATED SOLUTIONS}  

The original Bohr Hamiltonian \cite{Bohr} is
$$H = -{\hbar^2 \over 2B} \left[ {1\over \beta^4} {\partial \over \partial 
\beta} \beta^4 {\partial \over \partial \beta} + {1\over \beta^2 \sin 
3\gamma} {\partial \over \partial \gamma} \sin 3 \gamma {\partial \over 
\partial \gamma} \right. $$
\begin{equation}\label{eq:e1}
 \left. - {1\over 4 \beta^2} \sum_{k=1,2,3} {Q_k^2 \over \sin^2 
\left(\gamma - {2\over 3} \pi k\right) } \right] +V(\beta,\gamma),
\end{equation}
where $\beta$ and $\gamma$ are the usual collective coordinates describing the 
shape of the nuclear surface,
$Q_k$ ($k=1$, 2, 3) are the components of angular momentum, and $B$ is the 
mass parameter. 

It has been known for a long time \cite{Wilets} that exact separation of variables occurs in the corresponding Schr\"odinger equation if potentials
of the form $V(\beta,\gamma)=U(\beta)$ are used, i.e. if the potential 
is independent of the variable $\gamma$, thus corresponding to $\gamma$-soft
nuclei.  Then wavefunctions of the form  
$ \Psi(\beta,\gamma, \theta_i) = f(\beta) \Phi(\gamma, \theta_i)$
are used, where $\theta_i$ $(i=1,2,3)$ are the Euler angles describing 
the orientation of the nucleus in space. 

In the equation involving the angles, the eigenvalues of the second order 
Casimir operator of SO(5) occur, having the form 
 $\Lambda = \tau(\tau+3)$, where $\tau=0$, 1, 2, \dots is the quantum 
number characterizing the irreducible representations (irreps) of SO(5), 
called the ``seniority'' \cite{Rakavy}. This equation has been solved 
by B\`es \cite{Bes}.   

The values of angular momentum $L$ contained in each irrep of SO(5) 
(i.e. for each value of $\tau$) are given by the algorithm \cite{IA} 
$\tau=3\nu_\Delta +\lambda$, where $\nu_\Delta=0$, 1, \dots is the missing 
quantum number in the reduction SO(5) $\supset$ SO(3),  and 
$L=\lambda, \lambda+1, \ldots, 2\lambda-2, 2\lambda$ (with $2\lambda-1$ 
missing). The values of $L$ allowed for each $(\tau,\nu_\Delta)$ 
have been tabulated in \cite{IA,BonE5,CIE5}. 

The ``radial'' equation can be simplified by introducing \cite{IacE5} 
reduced energies $\epsilon = 
{2B\over \hbar^2} E$ and reduced potentials $u= {2B \over \hbar^2} U$.
The form of the solution of the radial equation depends on the choice made 
for $U(\beta)$. 

\subsection{E(5)} 

In the case of E(5) \cite{IacE5,CIE5}, a 5-dimensional (5-D) infinite well 
[$u(\beta) = 0$ if $\beta \leq \beta_W$, 
 $u(\beta)=\infty$  for $\beta > \beta_W$]  
is used, since the potential is expected to be flat at the point of a second 
order shape phase transition. Then the $\beta$-equation becomes a Bessel 
equation of order $\nu=\tau+3/2$, 
with eigenfunctions proportional to the Bessel functions 
$J_{\tau+3/2}(z)$ (with $z=\beta k$, $k =\sqrt{\epsilon}$), 
while the spectrum is determined by the zeros of the Bessel functions 
\begin{equation}\label{eq:e10}  
E_{\xi,\tau} = {\hbar^2 \over 2B} k^2_{\xi,\tau}, \qquad 
k_{\xi,\tau} = {x_{\xi,\tau} \over \beta_W}
\end{equation}
where $ x_{\xi,\tau}$ is the  $\xi$-th zero of the Bessel function 
$J_{\tau+3/2}(z)$. The spectrum is parameter free, up to an overall scale 
factor, which is fixed by normalizing the energies to the 
excitation energy of the first excited $2^+$ state, $2_1^+$. 
The $R_{4/2}=E(4_1^+)/E(2_1^+)$ ratio turns out to be 2.199~. 
The same holds for the $B(E2)$ values, which are normalized to the $B(E2)$ 
connecting the two lowest states, $B(E2; 2_1^+\to 0_1^+)$. 
The symmetry present in this case is \cite{CIE5}
E(5) $\supset$ SO(5) $\supset$ SO(3) $\supset$ SO(2). 

\subsection{Other solutions} 

For $u(\beta)= \beta^2/2$ one obtains the original solution of Bohr 
\cite{Bohr,Dussel}, which corresponds to a 5-D harmonic 
oscillator 
characterized by the symmetry U(5) $\supset$ SO(5) $\supset$ SO(3) $\supset 
$SO(2)
\cite{CM870}, the eigenfunctions being proportional to 
Laguerre polynomials \cite{Mosh1555},
and the spectrum having the simple form  
$ E_N = N+5/2$, with $ N=2\nu+\tau$, and $\nu=0,1,2,3,\ldots$, 
which has $R_{4/2}=2$. 
The spectra of the $u(\beta)=\beta^2/2$ potential and of the E(5) model 
become directly comparable by establishing the formal correspondence 
$\nu= \xi-1$. 

The Davidson potential $u(\beta)=\beta^2 + {\beta_0^4 \over \beta^2}$
(where $\beta_0$ is the position of the minimum of the potential)
\cite{Dav,Elliott,Rowe} also leads to eigenfunctions which are Laguerre
polynomials,  the energy eigenvalues being \cite{Elliott,Rowe}
(in $\hbar \omega=1$ units) 
\begin{equation}\label{eq:e7} 
E_{n,\tau} = 2n+1+ \left[ \left( \tau+{3\over 2} \right)^2 +\beta_0^4
\right]^{1/2} . 
\end{equation}
For $\beta_0=0$ the above mentioned original solution of Bohr [U(5)] 
is obtained, while for $\beta_0 \to \infty$ the O(6) limit of the IBM 
is obtained \cite{Elliott}. Therefore the Davidson potential provides 
a one-parameter bridge between U(5) and O(6). One can exploit this fact, 
by introducing a variational procedure \cite{varPLB,varPRC}, in which the 
rates of change of the $R_L=E(L_1^+)/E(2_1^+)$ energy ratios of the ground state 
band with respect to the parameter $\beta_0$ are maximized for each $L$ separately. The results lead to an energy spectrum very close to that of E(5)
\cite{varPLB}. The method has also been applied to other bands, as well as 
to $B(E2)$ transition rates \cite{varPRC}. 

The sequence of potentials $ u_{2n}(\beta) = {\beta^{2n} \over 2}$
(with $n$ being an integer) leads for $n=1$ to the Bohr case,  
while for $n \to\infty$ leads to the infinite well of E(5) 
\cite{Bender}.  Therefore this sequence of potentials provides a ``bridge''
between the U(5) symmetry and the E(5) model, using their common 
SO(5)$\supset$SO(3) chain of subalgebras for the classification of the 
spectra.  Solutions for $n\neq 1$ have been obtained numerically 
\cite{Ariasb4,Ariasb4b,BonE5}. 
The solutions for $n=2$, 3, 4, labelled as E(5)-$\beta^4$, E(5)-$\beta^6$, and 
E(5)-$\beta^8$, lead to $R_{4/2}=2.093$, 2.135, and 2.157 respectively. 
Complete level schemes have been given in Ref. \cite{BonE5}. 

A bridge complementary to the one just mentioned, i.e. a bridge spanning the 
region between E(5) and the $\gamma$-soft rotor O(5), has been obtained by using an infinite well 
potential with boundaries $\beta_M > \beta_m >0$ \cite{O5CBS}. The model, 
called O(5)-CBS, since it is a $\gamma$-soft analog of the confined $\beta$-soft 
(CBS) rotor model \cite{PG,DP}, contains one free parameter, 
$r_\beta= \beta_m/\beta_M$, the value $r_\beta=0$ corresponding to the E(5) model, and $r_\beta \to 1$ giving the $\gamma$-soft rotor [O(5)] limit.   

Other solutions obtained in this framework are listed below. 

a) A version of E(5) using a well of finite depth, instead of an infinite one, 
has been developed \cite{finitew}. 

b) The sextic oscillator, which is a quasi-exactly soluble \cite{Turb,Ushver} potential,
has also been used as a $\gamma$-independent potential \cite{sextic} 

c) Coulomb-like and Kratzer-like potentials have been used in Ref. \cite{FortE5}. 

d) A linear potential has been considered in Ref. \cite{Fortrev}, where a review 
of potentials used in this framework is given. 

e) A hybrid model employing a harmonic oscillator for $L\le 2$ and an infinite 
square well potential for $L\geq 4$ has been developed \cite{RadFae}. 

\subsection{Experimental manifestations of E(5)} 

The first nucleus to be identified as exhibiting E(5) behaviour was 
$^{134}$Ba \cite{CZE5}, while $^{102}$Pd \cite{Zamfir} also seems 
to provide a very good candidate. Further studies on $^{134}$Ba 
\cite{AriasE2} and $^{102}$Pd \cite{Har}, in which  
no backbending occurs in the ground state band, 
which remains in excellent agreement with the parameter-free E(5) predictions 
up to high angular momenta, reinforced this conclusion. 
$^{104}$Ru \cite{Frank}, $^{108}$Pd
\cite{Zhang}, $^{114}$Cd \cite{Long}, and $^{130}$Xe \cite{Liu} 
have also been suggested as possible candidates. A systematic search
\cite{ClarkE5,Kirson} on available data on energy levels and B(E2) transition 
rates suggested $^{102}$Pd, $^{106,108}$Cd, $^{124}$Te, $^{128}$Xe, and 
$^{134}$Ba as possible candidates, singling out $^{128}$Xe as the best one, 
in addition to $^{134}$Ba. 
This is in agreement with a recent report \cite{Kneissl}
on measurements of E1 and M1 strengths of $^{124-136}$Xe carried out at 
Stuttgart, which provides evidence for a shape phase transition around 
$A\simeq130$. 
$^{128}$Xe has been measured (November 2006) in  Jyv\"askyl\"a \cite{Hariss}.  
Recently, $^{58}$Cr has also been suggested as a candidate \cite{Marginean}.  

The assumption of a flat $\beta$-potential in the E(5) symmetry has been tested 
by constructing potential energy surfaces (PESs) for nuclei close to the E(5) symmetry, 
through the use of relativistic mean field theory \cite{FossionI}. It has been found that the relevant PESs come out quite flat, corroborating the E(5) 
assumption. 

\subsection{Odd nuclei: E(5/4) and E(5/12)} 

The models discussed so far are appropriate for even--even nuclei. Odd nuclei 
can be treated by coupling E(5), describing the even--even part of an odd nucleus, to the odd nucleon by the five-dimensional spin--orbit interaction
\cite{IacE54,CIE5}. If the odd nucleon is in a $j=3/2$ level, the 
E(5/4) model \cite{IacE54,CIE5} occurs, while if the odd nucleon lives in a
system of levels with $j=1/2$, 3/2, 5/2, the E(5/12) model \cite{E512,E512b} 
is obtained. Shape phase transitions from spherical to $\gamma$-unstable 
shapes in odd nuclei have also been considered \cite{AlonsoI,AlonsoII} in the 
framework
of the Interacting Boson Fermion Model \cite{IVI,FVI} for the case of an odd 
nucleon in a $j=3/2$ level. Shape phase transitions in odd nuclei have also 
been considered \cite{Jolie70} for the case of an odd nucleon in a 
system of levels with $j=1/2$, 3/2, 5/2 
in the framework of the U(5/12) supersymmetry  \cite{IVI,FVI}, giving good 
results in the Os--Hg region.  

A first effort to locate nuclei exhibiting the E(5/4) symmetry has been carried 
out for $^{135}$Ba \cite{Fetea}, with mixed results. The Ir--Au region might be 
a more appropriate one, since the U(6/4) supersymmetry has been found there 
\cite{IVI,FVI}. $^{63}$Cu, despite its small size, could also be a good candidate for E(5/4), as discussed in Ref. \cite{IacE54}. 

\section{X(5) AND RELATED SOLUTIONS} 

In the case of X(5) one tries to solve the Bohr Hamiltonian of Eq. (\ref{eq:e1}) 
for potentials of the form $u(\beta,\gamma)=u(\beta)+u(\gamma)$, seeking 
solutions of the relevant Schr\"odinger equation having the form 
$ \Psi(\beta, \gamma, \theta_i)= \phi_K^L(\beta,\gamma) 
{\cal D}_{M,K}^L(\theta_i)$, 
where $\theta_i$ ($i=1$, 2, 3) are the Euler angles, ${\cal D}(\theta_i)$
denote Wigner functions of them, $L$ are the eigenvalues of angular momentum, 
while $M$ and $K$ are the eigenvalues of the projections of angular 
momentum on the laboratory-fixed $z$-axis and the body-fixed $z'$-axis 
respectively. One is interested in cases near axial symmetry, i.e. close 
to $\gamma=0$. Thus one uses a harmonic oscillator potential 
$u(\gamma)=(3c)^2 \gamma^2/2$. Near $\gamma=0$ the last term in the Bohr 
Hamiltonian can be rewritten in the form \cite{IacX5}
\begin{equation}\label{eq:e3} 
\sum _{k=1,2,3} {Q_k^2 \over \sin^2 \left( \gamma -{2\pi \over 3} k\right)}
\approx {4\over 3} (Q_1^2+Q_2^2+Q_3^2) +Q_3^2 \left( {1\over \sin^2\gamma}
-{4\over 3}\right).  
\end{equation}
Using this result in the Schr\"odinger equation corresponding to 
the Hamiltonian of Eq. (\ref{eq:e1}), introducing reduced energies 
 $\epsilon = 2B E /\hbar^2$ and reduced potentials $u = 2B V /\hbar^2$
as in the E(5) case,  
and taking into account that the reduced potential is of the form 
$u(\beta, \gamma) = u(\beta) + u(\gamma)$, the Schr\"odinger equation can 
be separated into two equations \cite{IacX5}. 

In the equation containing the $\gamma$-variable, $\beta^2$ denominators 
remain, which are replaced by their average values over the $\beta$ wavefunctions, $\langle \beta^2 \rangle$. Taking into account the simplifications imposed by $\gamma$ being close to zero, the relevant equation 
takes the form corresponding to a two-dimensional harmonic oscillator 
in $\gamma$, having wavefunctions proportional to Laguerre polynomials
\cite{IacX5}. 

The form of the solution of the radial equation depends on the choice made 
for $u(\beta)$. 

\subsection{X(5)}  

In X(5) \cite{IacX5}
a 5-D infinite well potential is used, as in E(5). 
The relevant equation is again a Bessel equation, but with order 
\begin{equation}\label{eq:e9a} 
\nu= \left( {L(L+1)\over 3} +{9\over 4}\right)^{1/2}.
\end{equation}
The solutions still have the form of Eq. (\ref{eq:e10}), with $(\xi,\tau)$ 
replaced by $(s,L)$, where $s$ is the order of the relevant root of the Bessel 
function $J_\nu(k_{s,L} \beta)$. 
The relevant exactly soluble model 
is labelled as X(5) (which is not meant as a group label, although 
there is relation to projective representations of E(5), the Euclidean 
group in 5 dimensions \cite{IacX5}). 
The total energy has the form 
\begin{equation}\label{eq:e10a}
E(s,L,n_\gamma,K,M)= E_0 + B (x_{s,L})^2 + A n_\gamma + C K^2, 
\end{equation}
where $n_\gamma$ is the quantum number of the two-dimensional oscillator 
occurring in the $\gamma$-equation, while $E_0$, $A$, $B$, $C$ are free 
parameters. From this equation it is clear that the spectra of the ground 
state and $\beta$ bands only depend on an arbitrary scale, fixed by normalizing them to the energy of the $2_1^+$ state, while the bandheads of the $\gamma$ bands are parameter dependent. (The spacings within the $\gamma$ bands are
however fixed \cite{Bijker}.)  The $R_{4/2}$ ratio is 2.904~. 

\subsection{Other solutions} 

For $u(\beta)= \beta^2/2$ one obtains an exactly soluble model which has 
been called X(5)-$\beta^2$ \cite{BonX5}, the eigenfunctions being proportional to Laguerre polynomials and the spectrum having the  form
\begin{equation}\label{eq:e13a} 
E_{n,L}= 2n+1 + \sqrt{ {9\over 4}+{ L(L+1)\over 3} }, 
\qquad n=0,1,2,\ldots 
\end{equation}
with $R_{4/2}=2.646$. 
The spectra of the $u(\beta)=\beta^2/2$ potential and of the X(5) model 
become directly comparable by establishing the formal correspondence 
$n= s-1$, where $n$ is the usual oscillator quantum number. 

The Davidson potential $u(\beta)=\beta^2 + {\beta_0^4 \over \beta^2}$
(where $\beta_0$ is the position of the minimum of the potential)
\cite{Dav,Elliott,Rowe} also leads to eigenfunctions which are Laguerre
polynomials,  the energy eigenvalues being \cite{varPLB,varPRC}
(in $\hbar \omega=1$ units) 
\begin{equation}\label{eq:e17} 
E_{n,L} =
2n+1+ \left[ {1\over 3} L(L+1) + {9\over 4} +\beta_0^4
\right]^{1/2} . 
\end{equation}
For $\beta_0=0$ the above mentioned X(5)-$\beta^2$ solution 
is obtained, while for $\beta_0 \to \infty$ the rigid rotor limit 
is obtained. Therefore the Davidson potential provides 
a one-parameter bridge between X(5)-$\beta^2$ and the rigid rotor. 
One can exploit this fact, 
by introducing a variational procedure \cite{varPLB,varPRC}, in which the 
rates of change of the $R_L=E(L_1^+)/E(2_1^+)$ energy ratios of the ground state 
band with respect to the parameter $\beta_0$ are maximized for each $L$ separately. The results lead to an energy spectrum very close to that of X(5)
\cite{varPLB}. The method has also been applied to other bands, as well as 
to $B(E2)$ transition rates \cite{varPRC}. 

The sequence of potentials 
$ u_{2n}(\beta) = {\beta^{2n} \over 2}$  
(with $n$ being an integer) leads for $n=1$ to the X(5)-$\beta^2$ case,  
while for $n \to\infty$ leads to the infinite well of X(5) 
\cite{Bender}.  Therefore this sequence of potentials provides a ``bridge''
between the X(5)-$\beta^2$ solution and the X(5) model, 
in the region lying between U(5) and X(5).  
 Solutions for $n\neq 1$ have been obtained numerically \cite{BonX5}. 
The solutions for $n=2$, 3, 4, labelled as X(5)-$\beta^4$, X(5)-$\beta^6$, and 
X(5)-$\beta^8$, lead to $R_{4/2}=2.769$, 2.824, and 2.852 respectively. 
Complete level schemes have been given in Ref. \cite{BonX5}.

A bridge complementary to the one just mentioned, i.e. a bridge spanning the 
region between X(5) and the rigid rotor, has been obtained by using an infinite well 
potential with boundaries $\beta_M > \beta_m >0$ \cite{PG,DP}. The model, 
called the confined $\beta$-soft 
(CBS) rotor model \cite{PG,DP}, contains one free parameter, 
$r_\beta= \beta_m/\beta_M$, the value $r_\beta=0$ corresponding to the X(5) model, and $r_\beta \to 1$ giving the rigid rotor limit.   

Other solutions obtained in this framework are listed below. 

a) A  potential with linear sloped walls has been considered in Ref. \cite{sloped}. The sloped walls result in a slower increase of the energy levels of the $\beta$ band as a function of $L$ in this model, as compared to X(5). This feature improves agreement to experiment. 

b) Coulomb-like and Kratzer-like potentials have been used in Ref. 
\cite{FortX5}. 

c) The approximate separation of variables used in X(5) has been tested recently 
through exact numerical diagonalization of the Bohr Hamiltonian 
\cite{Caprio72}, using a recently introduced \cite{Rowe735,Rowe45,Rowe753}
computationally tractable version of the Bohr--Mottelson collective model.

d) Exact separation of variables can be achieved by using potentials of 
the form $u(\beta,\gamma)=u(\beta)+u(\gamma)/\beta^2$ \cite{Wilets}.
This possibility has been recently exploited for the construction of 
exactly separable (ES) analogues of the X(5) and X(5)-$\beta^2$ models, 
labelled as ES-X(5) and ES-X(5)-$\beta^2$ respectively \cite{ESX5}, 
as well as for the construction of ES-D \cite{ESD}, the  exactly separable 
version of the Bohr Hamiltonian with a Davidson potential as $u$($\beta$) 
and a stiff harmonic oscillator for
$u$($\gamma$) centered at $\gamma=0^{\circ}$. In this
model, called exactly separable Davidson (ES-D), the ground state
band, $\gamma$ band and $0_2^+$ band are all treated on an equal
footing \cite{IacCam}. The bandheads, energy spacings within bands, and 
a number of interband and intraband $B(E2)$ transition rates are well
reproduced for almost all well-deformed rare earth and actinide
nuclei using two parameters ($\beta_{0}$, $\gamma$ stiffness).
Insights regarding the recently found correlation between $\gamma$
stiffness and the $\gamma$-bandhead energy \cite{stiff}, as well as the long
standing problem of producing a level scheme with Interacting
Boson Approximation SU(3) degeneracies from the Bohr Hamiltonian,
have also been obtained.

e) The use of periodic $u(\gamma)$ potentials in the X(5) framework 
has been recently considered in Ref. \cite{Gheorghe}.  

A review of potentials used in this framework is given in Ref. \cite{Fortrev}. 

\subsection{X(3)}  

The special case in which $\gamma$ is frozen to $\gamma=0$, while 
an infinite square well potential is used in $\beta$, leads to an exactly 
separable three-dimensional model, which has been called X(3) \cite{X3}. 
This model involves three variables, $\beta$ and the two angles used in 
spherical coordinates, since the condition $\gamma =0$ guarantees an axially
symmetric prolate shape, for which the two angles of the spherical coordinates suffice for determining its orientation in space. 

Exact separation of variables is possible in this case. The equation involving 
the angles has the usual spherical harmonics as eigenfunctions, the relevant 
eigenvalues being $L(L+1)$, while the $\beta$-equation, in which an infinite 
square well potential is used, takes the form of a Bessel equation. The radial solutions have the same form as in X(5), but with order 
\begin{equation}\label{eq:e13c} 
\nu=\sqrt{\frac{L(L+1)}{3}+\frac{1}{4}}, 
\end{equation}
which should be compared to Eq. (\ref{eq:e9a}). 
It should be noticed that in E(3), the 
Euclidean algebra in 3 dimensions, which is the semidirect sum
of the T$_3$ algebra of translations in 3 dimensions and the SO(3) algebra of 
rotations in 3 dimensions \cite{Barut}, the eigenvalue equation of the square 
of the total momentum, which is a second-order Casimir operator of the 
algebra, also leads \cite{Barut,BonE5} to a similar solution, but with 
$\nu=L+{1\over 2}=\sqrt{L(L+1)+{1\over 4}}$.    

From the symmetry of the wave functions with respect 
to the plane which is orthogonal to the symmetry axis of the nucleus and goes 
through its center,
follows that the angular momentum $L$ can take only even nonnegative values.
Therefore no $\gamma$-bands appear in the model, as 
expected, since the $\gamma$ degree of freedom has been frozen. 
The $R_{4/2}$ ratio is 2.44~. 
Complete level schemes have been given in \cite{X3}. 

\subsection{Experimental manifestations of X(5)} 

The first nucleus to be identified as exhibiting X(5) behaviour was 
$^{152}$Sm \cite{CZX5}, followed by $^{150}$Nd \cite{Kruecken}. 
Further work on $^{152}$Sm \cite{ZamfirSm,Clark,CZK,Bijker} and $^{150}$Nd
\cite{Clark,CZK,Zhao} reinforced this conclusion. The neighbouring N=90 
isotones $^{154}$Gd \cite{Tonev,Dewald} and $^{156}$Dy \cite{Dewald,CaprioDy}
were also seen to provide good X(5) examples, the latter being of inferior
quality. In the heavier region, $^{162}$Yb \cite{McC162Yb} and 
$^{166}$Hf \cite{McC166Hf} have been considered as possible candidates.
More recent experiments on $^{176}$Os and $^{178}$Os \cite{DewaldOs}
indicate that the latter 
is a good example of X(5). 
A systematic study \cite{ClarkX5} 
of available experimental data on energy levels and B(E2) transition rates 
suggested $^{126}$Ba and $^{130}$Ce as possible good candidates, in 
addition to the N=90 isotones of Nd, Sm, Gd, and Dy. A similar study in 
lighter nuclei \cite{Brenner} suggested $^{76}$Sr, $^{78}$Sr and $^{80}$Zr 
as possible candidates.    
$^{104}$Mo has been suggested as a candidate for X(5) based on available spectra
\cite{BizMo,Brenner}, but later studies on $B(E2)$ values gave results close to the
rigid rotor limit \cite{Hutter}. 
$^{122}$Ba \cite{Fransen} is currently under consideration, since its ground 
state bands coincides with this of X(5). 
Recent measurements \cite{Balab}
on $^{128}$Ce indicate that this nucleus is  a good example of X(5).
This is expected, since $^{128}$Ce, having 8 valence protons and 12 valence 
neutron holes, matches $^{152}$Sm, possessing 12 valence protons and 8 
valence neutrons, which is a good example of X(5). 

The assumption of a flat $\beta$-potential in the X(5) symmetry has been tested 
by constructing potential energy surfaces (PESs) for nuclei close to the X(5) symmetry, 
through the use of relativistic mean field theory 
\cite{FossionI,Meng25,Sheng20,Yu15}. It has been found that the relevant PESs exhibit a bump in the middle, in accordance to 
calculations using an effective $\beta$ deformation, determined 
by variation after angular momentum projection and two-level mixing
\cite{LeviX5}, as well as in Nilsson-Strutinsky-BCS calculations 
\cite{ZhangSm} for $^{152}$Sm and $^{154}$Gd.

\section{Z(5) AND RELATED MODELS}  

Z(5) \cite{Z5} is an analogue of X(5) appropriate for triaxial nuclei. 
In both cases potentials of the form 
$u(\beta,\gamma)=u(\beta)+u(\gamma)$ are considered. 
In the X(5) case the Hamiltonian is simplified by focusing 
attention near $\gamma=0$, which corresponds to prolate axially symmetric nuclei. In Z(5) attention is focused near $\gamma=\pi/6$,
which corresponds to maximally triaxial shapes. It is known 
\cite{MtVNPA} that for $\gamma=\pi/6$ the projection of the 
angular momentum on the body-fixed $\hat x'$-axis, labelled
as $\alpha$, is a good 
quantum number, while the projection on the body-fixed $\hat z'$-axis,
labelled as $K$, is not a good quantum number. One then seeks
solutions of the relevant Schr\"odinger equation having the form 
$ \Psi(\beta, \gamma, \theta_i)= \phi_\alpha^L(\beta,\gamma) 
{\cal D}_{M,\alpha}^L(\theta_i)$, 
where $\theta_i$ ($i=1$, 2, 3) are the Euler angles, ${\cal D}(\theta_i)$
denote Wigner functions of them, $L$ are the eigenvalues of angular momentum, 
while $M$ and $\alpha$ are the eigenvalues of the projections of angular 
momentum on the laboratory-fixed $z$-axis and the body-fixed $x'$-axis 
respectively. 
$\alpha$ has to be an even integer \cite{MtVNPA}.
Instead of the projection $\alpha$ of the angular momentum on the 
$\hat x'$-axis, it is customary to introduce the wobbling quantum number 
\cite{MtVNPA,BM} $n_w=L-\alpha$. 

One is interested in cases near maximal triaxiality, i.e. close 
to $\gamma=\pi/6$. Thus one uses a harmonic oscillator potential 
$ u(\gamma)= c \left(\gamma-{\pi\over 6}\right)^2 /2=  c 
{\tilde \gamma}^2/2$, with $\tilde \gamma = \gamma -\pi/ 6$. 

In the case in which the potential 
has a minimum around $\gamma =\pi/6$ one can write  the last term of Eq. 
(\ref{eq:e1}) in the form 
\begin{equation}\label{eq:e2a} 
\sum _{k=1,2,3} {Q_k^2 \over \sin^2 \left( \gamma -{2\pi \over 3} k\right)}
\approx Q_1^2+4 (Q_2^2+Q_3^2) = 4(Q_1^2+Q_2^2+Q_3^2)-3Q_1^2. 
\end{equation}
Using this result in the Schr\"odinger equation corresponding to 
the Hamiltonian of Eq. (\ref{eq:e1}), introducing \cite{IacX5} reduced energies
 $\epsilon = 2B E /\hbar^2$ and reduced potentials $u = 2B V /\hbar^2$,  
and assuming \cite{IacX5} that the reduced potential can be separated into two 
terms, one depending on $\beta$ and the other depending on $\gamma$, i.e. 
$u(\beta, \gamma) = u(\beta) + u(\gamma)$, the Schr\"odinger equation can 
be separated into two equations \cite{Z5}. 

In the equation containing the $\gamma$-variable, $\beta^2$ denominators 
remain, which are replaced, in analogy with X(5), by their average values over the $\beta$ wavefunctions, $\langle \beta^2 \rangle$. 
Taking into account the simplifications imposed by $\gamma$ being close to $\pi/6$, the relevant equation 
takes the form corresponding to a simple one-dimensional harmonic oscillator 
in $\gamma$, having wavefunctions proportional to Hermite polynomials
\cite{Z5}. 

The form of the solution of the radial equation depends on the choice made 
for $u(\beta)$. 

\subsection{Z(5)} 

In Z(5) a 5-D infinite well potential is used, as in X(5). 
The relevant equation is again a Bessel equation, but with order  
\begin{equation}\label{eq:e9b} 
\nu = {\sqrt{4L(L+1)-3\alpha^2+9}\over 2}= 
{\sqrt{L(L+4)+3n_w(2L-n_w)+9}\over 2}.   
\end{equation}
The solutions still have the form of Eq. (\ref{eq:e10}), with $(\xi,\tau)$ 
replaced by $(s,\nu)=(s,n_W,L)$, where $s$ is the order of the relevant root of the Bessel 
function $J_\nu(k_{s,\nu} \beta)$. 
The relevant exactly soluble model 
is labelled as Z(5) (which is not meant as a group label).

The total energy has the form 
\begin{equation}\label{eq:e15} 
E(s,n_W,L,n_{\tilde\gamma})= E_0 + A (x_{s,\nu})^2 + B n_{\tilde \gamma}, 
\end{equation}
where $n_{\tilde\gamma}$ is the quantum number of the one-dimensional oscillator 
occurring in the $\gamma$-equation, while $E_0$, $A$, $B$ are free 
parameters.

The wobbling quantum number $n_w$ labels a series of bands 
with  $L=n_w,n_w+2,n_w+4, \dots$ (with $n_w > 0$) next to the ground state 
band (with $n_w=0$) \cite{MtVNPA}.  
The ground state band corresponds to $s=1$, $n_w=0$ and has $R_{4/2}=2.350$~. 
We shall refer to the model corresponding to this solution as Z(5)
(which is not meant as a group label), in analogy to the E(5) \cite{IacE5}, 
and X(5) \cite{IacX5} models.  
Complete level schemes have been given in \cite{Z5,PetrCam}. 
A preliminary comparison to experiment has suggested $^{192-196}$Pt 
as possible Z(5) candidates \cite{Z5,PetrCam}. 

\subsection{Other solutions} 

Solutions in the vicinity of $\gamma=\pi/6$ have also been worked out 
considering potentials of the form $u(\beta,\gamma)= u(\beta)+ 
u(\gamma)/\beta^2$ \cite{Forttri,FortHey}, which allow for an exact
separation of variables. Coulomb, Kratzer, harmonic, Davidson, and 
infinite square well potentials have been considered as $u(\beta)$ 
in this approach \cite{Forttri,FortHey}, while a displaced harmonic oscillator 
has been used as $u(\gamma)$. A periodic potential 
$u(\gamma)= \mu/\sin^2(3\gamma)$ has also been considered \cite{DeBae}. 
A solution similar to Z(5), but with a $u(\gamma)$ proportional to 
$\cos^2(3\gamma)$, has also been considered \cite{Jolos}. 

\subsection{Z(4)} 

The special case in which $\gamma$ is frozen to $\gamma=\pi/6$, while 
an infinite square well potential is used in $\beta$, leads to an exactly 
separable four-dimensional model, which has been called Z(4) \cite{Z4}. 
This model involves four variables, $\beta$ and the three Euler angles,
since $\gamma$ in this model is not treated as a variable but as a parameter,
as in the model of Davydov and Chaban \cite{DavCha}. 

Exact separation of variables is possible in this case. The equation involving 
the Euler angles has been solved by Meyer-ter-Vehn \cite{MtVNPA}, the eigenfunctions being appropriate combinations of Wigner functions. 
The $\beta$-equation, in which an infinite 
square well potential is used, takes the form of a Bessel equation. The radial solutions have the same form as in Z(5), but with order 
\begin{equation}\label{eq:e48}
\nu=\sqrt{L(L+1) - \frac{3}{4}\,\alpha^2 + 1}
={\sqrt{L(L+4)+3 n_w(2L-n_w)+4}\over 2},
\end{equation}
which should be compared to Eq. (\ref{eq:e9b}). 
where the various symbols have the same meaning as in Z(5). 
The $R_{4/2}$ ratio is 2.226~. 
Complete level schemes have been given in \cite{Z4,PetrCam}. 
A preliminary comparison to experiment has suggested $^{128-132}$Xe as possible 
Z(4) candidates \cite{Z4,PetrCam}. 

\subsection{Transition from axial to triaxial shapes} 

A special solution of the Bohr Hamiltonian corresponding to a transition from axially deformed to triaxially deformed shapes has been given in Ref. \cite{IacY5}, called Y(5). The proton--neutron triaxiality occurring in the 
SU(3)$^*$ limit \cite{Dieperink} of IBM-2 \cite{IA} has triggered the detailed 
study of the phase structure of IBM-2 \cite{AriasPRL,CaprioPRL,CaprioAP}. 
The main features of proton--neutron triaxiality are a low-lying $K=2$ band 
and $B(E2)$s resembling closely these of the Davydov model \cite{Davydov}.   

\section{CONCLUSIONS} 

A great and still growing interest has been developed in the last five years
in special solutions of the Bohr Hamiltonian, in relation to shape phase 
transitions and critical point symmetries in nuclei.  
Extensions of these ideas in many directions are ongoing, including 
the consideration of dipole \cite{Kneissl} and octupole 
\cite{Bizzoct,BizzoctII,AQOA,piAQOA} degrees of freedom. 
Many developments relevant to shape phase transitions and critical point 
symmetries, not mentioned in this work (which is not a review article but rather 
a biased brief account of topics related to the authors' work), can be
traced from the references in \cite{Rowe759,LoBianco,CastenNP}.

\end{document}